\documentclass[12pt]{iopart}

\usepackage{soul,xcolor}
\setstcolor{red}

\usepackage{graphicx}% Include figure files
\usepackage{dcolumn}% Align table columns on decimal point
\usepackage{bm}% bold math
\usepackage{color}

\begin{document}
\def \beq{\begin{equation}}
\def \eeq{\end{equation}}
\def \bea{\begin{eqnarray}}
\def \eea{\end{eqnarray}}
\def \bem{\begin{displaymath}}
\def \eem{\end{displaymath}}
\def \P{\Psi}
\def \Pd{|\Psi(\boldsymbol{r})|}
\def \Pds{|\Psi^{\ast}(\boldsymbol{r})|}
\def \Po{\overline{\Psi}}
\def \bs{\boldsymbol}
\def \bl{\bar{\boldsymbol{l}}}
%%%%%%%%%%%%%%%%%%%%%%%%%%%%%%%%%%%%%%%%%%%%%%%%%%
\title{Integrable nonlocal asymptotic reductions of physically significant nonlinear equations}
\author{Mark J. Ablowitz$^1$ and Ziad H. Musslimani$^2$}
\address{$^1$Department of Applied Mathematics, University of Colorado, Campus Box 526,  Boulder, Colorado 80309-0526\\
$^2$Department of Mathematics, Florida State University, Tallahassee, FL 32306-4510}
\date{\today}
%%%%%%%%%%%%%%%%%%%%%%%%%%%%%%%%%%%%%%%%%%%%%%%%%
%%%%%%%%%%%%%%%%%%%%%%%%%%%%%%%%%%%%%%%%%%%%%%%%%
\begin{abstract}
%%%%%%%%%%%%%%%%%%%%%%%%%%%%%%%%%%%%%%%%%%%%%%%%%
%%%%%%%%%%%%%%%%%%%%%%%%%%%%%%%%%%%%%%%%%%%%%%%%%
Quasi-monochromatic {\it complex} reductions of a number of physically important equations are 
obtained. Starting from the cubic nonlinear Klein-Gordon (NLKG), the Korteweg-deVries (KdV) and water wave equations, it is shown that the leading order asymptotic approximation can be transformed to the well-known integrable AKNS system \cite{AKNS} associated with second order (in space) nonlinear wave equations. This in turn establishes, for the first time, an important physical connection between the recently discovered nonlocal integrable reductions of the AKNS system
and physically interesting equations. Reductions include the parity-time, reverse space-time and reverse time nonlocal nonlinear Schr\"odinger equations. 
%%%%%%%%%%%%%%%%%%%%%%%%%%%%%%%%%%%%%%%%%%%%%%%%%
%%%%%%%%%%%%%%%%%%%%%%%%%%%%%%%%%%%%%%%%%%%%%%%%%
\end{abstract}
%%%%%%%%%%%%%%%%%%%%%%%%%%%%%%%%%%%%%%%%%%%%%%%%%
\maketitle
%%%%%%%%%%%%%%%%%%%%%%%%%%%%%%%%%%%%%%%%%%%%%%%%%
\section{Introduction} 
Ever since the seminal work establishing that the KdV, nonlinear 
Schr\"odinger (NLS), sine-Gordon equations and many others are integrable 
\cite{KDV, BenneyNewell67, Bour, GGKM, ZS, AKNS}, there has been an enormous effort directed at finding solutions and understanding the mathematical and physical properties of these equations cf. \cite{Ablowitz2,Ablowitz1,Cal-Deg,NOVIKOV,Ablowitz3,Yang_book}. Methods include the Inverse Scattering Transform (IST) and direct methods such as Hirota, Darboux and B\"acklund transformations \cite{Hirota, Matveev, Rogers} amongst many others.
%%%%%%%%%%%%%%%%%%%%%%%%%%%%%%%%%%%%%%%%%%%%%%%%%
From a physics point of view, the classical NLS equation (given in normalized units)
\begin{equation}
iq_t=q_{xx} - 2\sigma q^2q^*, \sigma= \mp 1\;,
\label{CNLS}
\end{equation}
($q^*$ is the complex conjugate of $q$) plays a crucial role in modeling physical systems ranging from photonics to Bose-Einstein condensation to deep water fluid dynamics to mention a few \cite{Agrawal_1,Agrawal_2,BEC,Ablowitz4}. 
This NLS equation is universal, namely, it  arises generically as the slowly varying wave envelope approximation of a uniform wave train solution of the underlying governing equation. This is frequently referred to as the quasi-monochromatic approximation cf. \cite{Ablowitz4}. Importantly, this equation was deduced in \cite{AKNS} as a special symmetry reduction of a coupled system of evolution equations given by
%%%%%%%%%%%%%%%
\begin{equation}
iq_t=q_{xx} -2q^2r \;,
\label{NLq}
\end{equation}
\begin{equation}
-ir_t=r_{xx}-2r^2q \;,
\label{NLr}
\end{equation}
where $q(x,t)$ and $r(x,t)$ are potentials of the well-known AKNS $2\times2$ linear scattering 
problem \cite{AKNS}. Indeed, when $r=\sigma q^*$, one recovers Eq.~(\ref{CNLS}).\\
 %%%%%%%%%%%%%%%%%%%%%%%%%%%%%%%%%%%%%%%%%%%%%%%%%
For a few decades, it was thought that $r=\sigma q^*$ is the only interesting reduction of the AKNS scattering problem. Surprisingly, in 2013 Ablowitz and Musslimani \cite{AblowitzMusslimani} showed that there was another interesting reduction given by $r(x,t)=\sigma q^*(-x,t)$ giving rise to the integrable nonlocal NLS equation
%%%%%%%%%%%%%%%%%%%%%%%%%%%%%%%%
\begin{equation}
iq_t(x,t)=q_{xx}(x,t) - 2\sigma q^2(x,t)q^*(-x,t)\;.
\label{PTNLS}
\end{equation}
Motivated by studies of parity-time ($PT$) symmetry in quantum physics and 
optics cf. \cite{Bender, RKDM2, RKDM3,RKDM4, KYZ, DNCY,KGM,NatPhys}, 
they termed it the $PT$ symmetric NLS equation, (or PTNLS
for short). The  IST theory associated with PTNLS equation was studied in detail in  \cite{AblowitzMusslimani3}. Soon afterwards, Ablowitz and Musslimani found that there were yet two more reductions of the AKNS scattering problem leading to interesting nonlocal NLS type equations: These are $r(x,t)=\sigma q(-x,-t)$ and $r(x,t)=\sigma q(x,-t)$ giving rise to the so-called reverse space-time NLS (RSTNLS) and reverse time NLS (RTNLS) equations, respectively given by \cite{AblowitzMusslimani4}.
%%%%%%%%%%%%%%%%%%%%
\begin{equation}
iq_t(x,t)=q_{xx}(x,t) - 2 \sigma q^2(x,t)q(-x,-t),
\label{RSTNLS}
\end{equation}
%%%%%%%%%%%%%%%%%%%%%
\begin{equation}
iq_t(x,t)=q_{xx}(x,t) - 2 \sigma q^2(x,t)q(x,-t).
\label{RTNLS}
\end{equation}
%%%%%%%%%%%%%%%%%%%%%%%%%%%%%%%%%%%%%%%%%%%%%%%%%
These findings have ignited great interest in the emerging field of nonlocal $PT$ symmetric integrable systems focusing on its mathematical structure and physical properties
\cite{LouHuang, Yang1, Yang2, Yang3, Yang4, Xu, Wen, AFokas, Ma, mag, AblowitzLuoMusslimani1, AblowitzLuoMusslimani2, AblowitzLuoMusslimani3, AblowitzLuoMusslimani4, Chen1, Li, MG1, MG2, Ji,Yang_Chen,Rao}.
A natural question is whether and how such reductions can come out of physically interesting nonlinear wave equations, thus establishing connections between nonlocal integrable reductions and physically relevant models. In this paper, we address this very important issue and show that all of the above equations, i.e., the PTNLS, RSTNLS and RTNLS can be obtained as asymptotic quasi-monochromatic reductions from the nonlinear Klein-Gordon equation with a cubic nonlinear term, the KdV equation which has a quadratically nonlinear term and the more complicated nonlinear water wave equations. 
To do this, we modify the standard assumption of the form of the leading order solution, allowing it to be {\it complex.} This results in the remarkable observation that one gets a system of equations that is transformable to the standard AKNS $q,r$ system for second order spatial systems. Furthermore this observation connects the integrable KdV equation with all the above NLS type integrable flows.\\
%%%%%%%%%%%%%%%%%%%%%%%%%%%%%%%%%%%%%%%%%%%%%%%%%
%%%%%%%%%%%%%%%%%%%%%%%%%%%%%%%%%%%%%%%%%%%%%%%%%
\section{Nonlinear Klein-Gordon reductions} We begin by considering the NLKG equation (sometimes called the $\phi^4$ model in physics) 
\begin{equation}
\phi_{tt}-\phi_{xx}+\phi-\sigma \phi^3=0 \;,
\label{KG}
\end{equation}
which, for $\sigma=-1,$ can be obtained as an approximation to the sine-Gordon equation $\phi_{tt}-\phi_{xx}+\sin \phi=0$ by keeping the first two terms in the expansion of $\sin \phi$ for small $\phi$. Here and below we consider $x,t$ to be real.
We look for a ``small" quasi-monochromatic solution to Eq.~(\ref{KG}) in the form 
%%%%%%%%%%%%%%%%%%%
\begin{equation}
\label{multi-scale}
\phi=\phi(\theta, X,T;\epsilon), ~~\theta=k x-\omega t, X=\epsilon x, T=\epsilon t\;,
\end{equation}
%%%%%%%%%%%%%%%%%%
with the linear dispersion relation and group velocity given by $\omega^2-k^2=1, ~~\omega'(k)
=\frac{k}{\omega}.$ Substituting Eq.~(\ref{multi-scale}) in (\ref{KG}), expanding $\phi$ in a series 
in $\epsilon$, i.e., $\phi= \epsilon \phi_0+\epsilon^2 \phi_1+\epsilon^3 \phi_2+ \cdots$,
and setting all coefficients to zero yields a sequence of equations at $O(\epsilon^j), j=0,1,2...$
%%%%%%%%%%%%%%%%%%%%%%%%%%%%%%%%%%%%%%%%%%%%
%%%%%%%%%%%
\begin{equation}
L\phi_0 \equiv  (\partial_{\theta}^2+1)\phi_0 = 0 \;,
\label{E0}
\end{equation}
%%%%%%%%%%%
\begin{equation}
L\phi_1=  (2k\partial_{\theta}\partial_X+2\omega\partial_{\theta}\partial_T)\phi_0 \;,
\label{E1}
\end{equation}
%%%%%%%%%%%
\begin{equation}
L\phi_2=  (2k\partial_{\theta}\partial_X+2\omega\partial_{\theta}\partial_T)\phi_1-(\partial_T^2-\partial_X^2)\phi_0+\sigma \phi_0^3 \;.
\label{E2}
\end{equation}
%%%%%%%%%%%
The general solution to the leading order equation (\ref{E0}) is
\begin{equation}
\phi_0= A(X,T)e^{i\theta}+B(X,T)e^{-i\theta} \;.
\label{S0}
\end{equation}
Note that here we {\it do not} necessarily require $B=A^*$. Hence these solution are generally 
{\it complex} by considering complex initial data. We point out that complexifying physically interesting equations either through its independent variables (e.g. Painlev\'e type equations \cite{AF})
or by allowing the dependent variables (wave functions) to become complex valued \cite{SDYM} has been the subject of great interest in recent years  \cite{Fokas_complex1, Fokas_complex2, Bender_complex1, Bender_complex2,Bender_complex3} . Doing so leads to a range of novel 
phenomena \cite{kevrekidis}. 
%%%%%%%%%%%%%%%%%%%%%%%%%%%%%%%%%%%%%%%%%%%%%%
Substituting the solution given by (\ref{S0})  into  (\ref{E1}) leads to
$L\phi_1=  2i(k\partial_XA+\omega\partial_TA)e^{i\theta}-2i(k\partial_XB
+\omega\partial_TB)e^{-i\theta}.$
To remove secular terms we take 
\begin{equation}
2i(k\partial_XA+\omega\partial_TA)=\epsilon g_1+\epsilon^ 2g_2+\cdots \;,
\label{Sec1}
\end{equation}
\begin{equation}
-2i(k\partial_XB+\omega\partial_TB)=\epsilon h_1+\epsilon^ 2h_2+\cdots \;,
\label{Sec2}
\end{equation}
and require $\phi_1=0$ (all homogeneous solutions are incorporated into the leading order). The terms $g_1,g_2,..., h_1,h_2,...$ are determined by removal of secular terms at higher order. In this way the appropriate higher order nonlinear equations are obtained.
The equation at $O(\epsilon^3)$ is now given by
\begin{equation}
L\phi_2=  -(\partial_T^2-\partial_X^2)\phi_0+\sigma\phi_0^3+g_1e^{i\theta}+h_1e^{-i\theta} \;.
\label{PS2}
\end{equation}
Substituting the solution for $\phi_0$ from equation (\ref{S0}) determines $g_1,h_1$ to be
\begin{equation}
g_1=(\partial_T^2-\partial_X^2)A-3\sigma A^2B \;,
\label{Sg1}
\end{equation}
\begin{equation}
h_1=(\partial_T^2-\partial_X^2)B+3\sigma A^2B \;.
\label{Sh1}
\end{equation}
The remaining terms in equation (\ref{PS2}) satisfy 
\begin{equation}
L\phi_2= \sigma (A^3e^{3i\theta}+B^3e^{-3i\theta}) \;,
\label{PS2R}
\end{equation}
whose solution is given by
\begin{equation}
\phi_2= -\frac{\sigma}{8} (A^3e^{3i\theta}+B^3e^{-3i\theta}) \;.
\label{PS2RS}
\end{equation}
%%%%%%%%%%%%%%%%%%%%%%%%%%%%%%%%%%%%%%%%%%%%%%%
Using $g_1,h_1$ from equations (\ref{Sec1})-(\ref{Sec2}) we find the nonlinear equations
\begin{equation}
2i(\omega\partial_TA+k\partial_XA)-\epsilon ((\partial_T^2-\partial_X^2)A+3\sigma A^2B)=0 \;,
\label{NL1}
\end{equation}
\begin{equation}
-2i(\omega\partial_TB+k\partial_XB)-\epsilon ((\partial_T^2-\partial_X^2)B+3\sigma B^2A)=0 \;.
\label{NL2}
\end{equation}
The above equations are coupled NLS equations which we will transform to a more convenient form. Using the dispersion relation and the transformations $\xi=x-\omega'(k)T, ~\tau=\epsilon T,$ the nonlinear equation (\ref{NL1})-(\ref{NL2}) now read
\begin{equation}
i\partial_{\tau}A+\frac{\omega''(k)}{2}\partial_{\xi}^2A+\frac{3}{2\omega}\sigma A^2B=0 \;,
\label{NL1F}
\end{equation}
\begin{equation}
-i\partial_{\tau}B+\frac{\omega''(k)}{2}\partial_{\xi}^2B+\frac{3}{2\omega}\sigma B^2A=0 \;,
\label{NL2F}
\end{equation}
where we used $\omega''(k)=(1-\omega'^2)/\omega=1/\omega^3.$ By rescaling 
$\xi=x, \tau=\gamma t, A=aq,B=br$ with  $\gamma=2\omega^3, 3\sigma\omega^2ab=-2,$
 these equations transform to the $q,r$ system (\ref{NLq}) and (\ref{NLr}). These equations are the integrable AKNS pair that leads to NLS and its reductions cf. \cite{Ablowitz3,AblowitzMusslimani4}. In particular, as stated in the introduction, Eqns.~ (\ref{NLq}) and (\ref{NLr}) have exact reductions to the classical NLS (\ref{CNLS}) when $r=\sigma q^*$, the PTNLS (\ref{PTNLS}) when 
 $r(x,t)=\sigma q^*(-x,t)$ ({\ref{RSTNLS}) when $r(x,t)=\sigma q(-x,-t)$ and the RTNLS equation ({\ref{RTNLS}) when $r(x,t)=\sigma q(x,-t)$ cf. \cite{Ablowitz3}. Solutions to the nonlocal equations (\ref{PTNLS}) - (\ref{RTNLS}) were analyzed for decaying data in \cite{Ablowitz3,AblowitzMusslimani4} and for non decaying initial conditions in \cite{AblowitzLuoMusslimani1,AblowitzLuoMusslimani2}.\\
%%%%%%%%%%%%%%%%%%%%%%%%%%%%%%%%%%%%%%%%%%%%%%%%
%%%%%%%%%%%%%%%%%%%%%%%%%%%%%%%%%%%%%%%%%%%%%%%%
\section{Korteweg-deVries reductions} In this section we will consider the following normalized KdV equation
\begin{equation}
u_t+6uu_x+u_{xxx}=0 \;,
\label{KdV}
\end{equation}
to derive the $q,r$ system. Similar to the nonlinear KG case, we will look for a ``small" quasi-monochromatic solution to the KdV in the form given by Eq.~(\ref{multi-scale}), with the linear dispersion relation and group velocity given 
by $\omega=-k^3, ~~\omega'(k)=-3k^2.$ Expanding $u$ in a series in $\epsilon$: $u=\epsilon u_0+\epsilon^2 u_1+\epsilon^3 u_2 +\cdots$ yields a sequence of equations, of which the first three at $O(\epsilon^j), j=1,2,3$, are given by 
%%%%%%%%%%%%%%%%%%%%%%%%
\begin{equation}
\hat{L}u_0 \equiv \partial_{\theta}(\partial_{\theta}^2+1)u_0 = 0 \;,
\label{u_0}
\end{equation}
%%%%%%%%%%%%%%%%%%%%%%%%
\begin{equation}
\hat{L}u_1= -\partial_Tu_0-6ku_0\partial_{\theta}u_{0}-3k^2\partial_{\theta}^2\partial_{X}u_{0} \;,
%\partial_{\theta}^2\partial_Xu_0
\label{u_1}
\end{equation}
%%%%%%%%%%%%%%%%%%%%%%%%%%
\begin{eqnarray}
\hat{L}u_2&=&-\partial_Tu_1-6ku_0\partial_{\theta}u_{1}-6u_0\partial_{X}u_{0}
\nonumber \\
&-&6ku_1\partial_{\theta}u_{0}-3k^2\partial_{\theta}^2\partial_{X}u_1
-3k\partial_{\theta}\partial_{X}^2u_0 \;.
\label{u_2}
\end{eqnarray}
%%%%%%%%%%%%%%%%%%%%%%%%%
The solution to the leading order equation is given by
\begin{equation}
u_0=A(X,T)e^{i\theta}+B(X,T)e^{-i\theta}+M(X,T) \;.
\label{SK0}
\end{equation}
Here, we note that the solution to the leading order KdV asymptotic series is different from that of the nonlinear KG by adding a required mean term $M(X,T)$. Substituting this solution into (\ref{u_1}) yields
%%%%%%%%%%%%%%%%%%%%%%%%%%%%%%%%%%%%%%%%
\begin{eqnarray}
\hat{L}u_1&=& -(A_T+\omega'(k)A_X+6ikMA)e^{i\theta}-(B_T+\omega'(k)B_X
\nonumber \\
&-&6ikMB)e^{-i\theta}-M_T-6ikA^2e^{2i\theta}
+6ikB^2e^{-2i\theta} \;.
\label{PSK1}
\end{eqnarray}
%%%%%%%%%%%%%%%%%%%%%%%%%%%%%%%%%%%%%%%%%
We remove secular terms by taking 
\begin{equation}
A_T+\omega'(k)A_X+6ikMA=\epsilon\hat{g}_1+\epsilon^2\hat{g}_2+\cdots \;,
\label{SecK1}
\end{equation}
\begin{equation}
B_T+\omega'(k)B_X-6ikMB=\epsilon\hat{h}_1+\epsilon^2\hat{h}_2+\cdots \;,
\label{SecK2}
\end{equation}
\begin{equation}
M_T=\epsilon\hat{f}_1+\epsilon^2\hat{f}_2+\cdots \;,
\label{SecK3}
\end{equation}
and the solution $u_1$ is given by
\begin{equation}
u_1=\alpha_1e^{2i\theta}+\beta_1e^{-2i\theta}, ~~\alpha_1=\frac{A^2}{k^2}, \beta_1=\frac{B^2}{k^2} \;.
\label{SK1}
\end{equation}
%%%%%%%%%%%%%%%%%%%%%%%%%%%%%%%%%%%%%%%%%%%%%%%%
The equation at $O(\epsilon^3)$ is therefore given by
\begin{eqnarray}
\hat{L}u_2 = R\;,
\end{eqnarray}
where $R$ (not given here due to size) depends on the amplitudes and their derivatives.
To remove secular terms, consequently we find
\begin{equation}
\hat{g}_1=-6i\frac{A^2B}{k}-3ikA_{XX}-6(AM)_X \;,
\end{equation}
\begin{equation}
\hat{h}_1=6i\frac{AB^2}{k}+3ikB_{XX}-6(BM)_X \;,
\end{equation}
\begin{equation}
\hat{f}_1=6(AB)_X-6MM_X \;,
\end{equation}
hence the equations (\ref{SecK1})-(\ref{SecK3}) yield
\begin{equation}
A_T+\omega'(k)A_X+6ikMA=\epsilon(-6i\frac{A^2B}{k}-3ikA_{XX}-6(AM)_X)
\label{NFK1}
\end{equation}
\begin{equation}
B_T+\omega'(k)B_X-6ikMB=\epsilon(6i\frac{AB^2}{k}+3ikB_{XX}-6(BM)_X)
\label{NFK2}
\end{equation}
\begin{equation}
M_T=\epsilon(-6(AB)_X-6MM_X) \;.
\label{NFK3}
\end{equation}
Employing traveling coordinates, these equations can be put in the form given by 
Eqns.~(\ref{NL1F}) and  (\ref{NL2F}) with the nonlinear coefficient $3\sigma/(2\omega)$ being 
replaced by $6/k (k \neq 0).$ 
Thus the integrable KdV equation is directly connected to this integrable $q,r$ system (\ref{NLq})-(\ref{NLr}) and its reductions: classical NLS, PTNLS, RSTNLS, RTNLS.\\
%%%%%%%%%%%%%%%%%%%%%%%%%%%%%%%%%%%%%%%%%%%%%%%%
%%%%%%%%%%%%%%%%%%%%%%%%%%%%%%%%%%%%%%%%%%%%%%%%
\section{Water wave reductions} In this section, we show how one can obtain the AKNS reduction starting from the classical water wave equations governing ideal incompressible surface gravity waves. To do so, we start from the one dimensional, deep water wave equations given by
\begin{equation}
\phi_{xx}+\phi_{zz} =0, \;,\;\; -\infty<z< \eta \;;\;\;\; \lim_{z\to -\infty} \phi_z =0 \;,
\label{WWInt}
\end{equation}
%%%%%%%%%%%%%%%%%%%%%%%%%%%%%%%%%%%%%%%%%%%%%%
\begin{equation}
\eta_t+\eta_x\phi_x=\phi_z, ~~on ~~z= \eta \;,
\label{WWKin}
\end{equation}
%%%%%%%%%%%%%%%%%%%%%%%%%%%%%%%%%%%%%%%%%%%%%%
\begin{equation}
\phi_t +\phi_x^2+\phi_z^2+g\eta=0, ~~on ~~z= \eta \;,
\label{WWBern}
\end{equation}
%%%%%%%%%%%%%%%%%%%%%%%%%%%%%%%%%%%%%%%%%%%%%
where $\eta$ and $\phi$ are the fluid free surface elevation and velocity potential respectively.
For simplicity, deep water is considered. We assume that the amplitude is small. To analyze this system one expands $\phi$ around the small amplitude free surface $z=\eta$ in equations (\ref{WWKin})-(\ref{WWBern}) as follows
\begin{equation}
\phi(x, \eta)=\phi(x,0)+\phi_z(x,0) \eta+\frac{1}{2}\phi_{zz}(x,0) \eta^2+\cdots
\label{WWexp}
\end{equation}
Then we expand $\phi, \eta$ in the following asymptotic series
\begin{eqnarray}
\phi &=& \epsilon (Ae^{i\theta+|k|z}+Be^{-i\theta+|k|z})
\nonumber \\
&+&\epsilon^2(A_2e^{2i\theta+2|k|z}+B_2e^{-2i\theta+2|k|z}+\tilde{\phi})+\cdots
\label{WWexp1}
\end{eqnarray}
\begin{equation}
\eta=\epsilon (Ce^{i\theta}+De^{-i\theta})+\epsilon^2(C_2e^{2i\theta}+D_2e^{-2i\theta}+\tilde{\eta})+\cdots
\label{WWexp11}
\end{equation}
where the coefficients of $\phi$: $A,B,A_2,B_2, \tilde{\phi}$ depend on slow variables $X=\epsilon x, Z=\epsilon z, T=\epsilon t$ while the coefficients of $\eta$: $C,D,C_2,D_2,\tilde{\eta}$ depend only on $X=\epsilon x, T=\epsilon t$; the rapid phase is, as usual, given by $\theta=kx-\omega t$. We substitute (\ref{WWexp1})-(\ref{WWexp11}) into the water wave equations (\ref{WWInt})-(\ref{WWKin}), cf. \cite{Ablowitz4}. The leading order problem shows that the dispersion relation  satisfies $\omega^2=g|k|,$ and $A=gC/(i\omega), B=-gD/(i\omega)$.
At higher order we calculate $A_2,B_2,C_2,D_2$ in terms of the first harmonics C,D and it is found that the mean term $\tilde{\eta}$ (like the KdV case) is small; we find: $C_2=  \frac{ik^2}{\omega}CA,  A_2=O(\epsilon), D_2=-\frac{ik^2}{\omega}BD,  ~~ B_2=O(\epsilon),
\tilde{\eta}=O(\epsilon), \partial_X\tilde{\phi}-\frac{2k|k|}{\omega}(AB)_X=O(\epsilon).$
Amplitude equations are obtained for the first harmonics.
In traveling wave coordinates the leading order slowly varying first harmonic amplitudes $(C,D)$ associated with the wave elevation $\eta$ satisfy the same equations as (\ref{NL1F}) and  (\ref{NL2F}) only with the nonlinear coefficient $3\sigma/(2\omega)$ being replaced by
 $n_2=-2k^2\omega$. Rescaling these equations yield the general $q,r$ system (\ref{NLq})-(\ref{NLr}).} \\
%%%%%%%%%%%%%%%%%%%%%%%%%%%%%%%%%%%%%%%%%%%%%%%%%%
%%%%%%%%%%%%%%%%%%%%%%%%%%%%%%%%%%%%%%%%%%%%%%%%%%
Some additional remarks are in order. The only differences between deep water and finite depth 
($h$)} are that the dispersion relation and nonlinear coefficient $n_2$ are different cf. \cite{Ablowitz2}. We also note that after inspecting the above reduction the quasi-monochromatic limit of {\it two-dimensional} shallow water waves with surface tension  in shallow water ($kh \ll 1$) will follow along the same lines (see also \cite{BR69,DS74,Ablowitz2,Ablowitz1}). 
%sees that the quasi-monochromatic limit of {\it two-dimensional} shallow water waves with surface tension  in shallow water ($kh \ll 1$) will satisfy, after rescaling, (see also \cite{BR69,DS74,Ablowitz2,Ablowitz1})
%%%%%%%%%%%%%%%%%%%%%%%%%%%%%%%
%\begin{equation}
%iC_{\tau}-\sigma C_{xx}+C_{yy}=\sigma C^2D+C\Phi_x \;,
%\label{WW2dC}
%\end{equation}
%%%%%%%%%%%%%%%%%%%%%%%%%%%%%%
%\begin{equation}
%-iD_{\tau}-\sigma D_{xx}+D_{yy}=\sigma D^2C+D\Phi_x \;,
%\label{WW2dD}
%\end{equation}
%%%%%%%%%%%%%%%%%%%%%%%%%%%%%%
%\begin{equation}
%\sigma \Phi_{xx}+\Phi_{yy}=-2(CD)_{xx}, ~~\sigma=\pm 1 \;.
%\label{WW2dMn}
%\end{equation}
%%%%%%%%%%%%%%%%%%%%%%%%%%%%%%
%where $\Phi$ is related to the mean of $\phi$. 
One expects to find a system closely related to AKNS systems associated with second order nonlinear 
wave equations in two space dimensions which has reductions to the integrable classical 
Davey-Stewartson equation and the $PT$ symmetric Davey-Stewartson equation cf. \cite{AblowitzMusslimani4}. However details of this calculation are outside the scope of this paper.
\\
%%%%%%%%%%%%%%%%%%%%%%%%%%%%%%%%%%%%%%%%%%%%%%%%
%%%%%%%%%%%%%%%%%%%%%%%%%%%%%%%%%%%%%%%%%%%%%%%%
\section{Conclusion} Quasi-monochromatic {\it complex} reductions of a cubic nonlinear Klein-Gordon,
the KdV and water waves equations are considered. It is found that the asymptotic reductions satisfy the well-known AKNS $``q,r"$ system (\ref{NLq}) and (\ref{NLr}) for second order in space integrable nonlinear wave equations. As such, all integrable nonlocal reductions, recently reported in the literature, are contained. This includes the PTNLS, reverse space-time and reverse time NLS equations.\\
%%%%%%%%%%%%%%%%%%%%%%%%%%%%%%%%%%%%%%%%%%%%%%%%
%%%%%%%%%%%%%%%%%%%%%%%%%%%%%%%%%%%%%%%%%%%%%%%%
{\it Acknowledgement.} MJA was partially supported by NSF under Grant No. DMS-1712793.\\
%%%%%%%%%%%%%%%%%%%%%%%%%%%%%%%%%%%%%%%%%%%%%%%%%
%%%%%%%%%%%%%%%%%%%%%%%%%%%%%%%%%%%%%%%%%%%%%%%%
%%%%%%%%%%%%%%%%%%%%%%%%%%%%%%%%%%%%%%%%%%%%%%%%

%%%%%%%%%%%%%%%%%%%%%%%%%%%%%%%%%%%%%%%%%%%%%%%%%
\end{document}